\documentclass[conference]{IEEEtran}
\ifCLASSINFOpdf
\else
\fi
\usepackage[]{algorithm2e}
\usepackage{amsmath}
\usepackage{amssymb}
\usepackage{epsfig}
\usepackage{epstopdf}
\usepackage{color}
\usepackage{multicol}
\usepackage{multirow}
\usepackage{mathtools}
\usepackage{lipsum}
\usepackage{algorithmic}
\usepackage{mdframed}
\usepackage{subfigure}
\newcommand{\Fap}{\mathcal{T}_{ap}}
\newcommand{\fp}{{fp}}

\newcommand{\Nfp}{N_{fp}}

\newcommand{\sx}{m}
\newcommand{\vx}{\mathbf{\sx}}
\newcommand{\vc}{\mathbf{c}}

\newcommand{\vxch}{\tilde{\vx}}

\newcommand{\nfloor}{{F}}
\newcommand{\NC}{{N_c}}
\newcommand{\Nap}{{N_{ap}}}
\newcommand{\defeq}{\triangleq}

\newcommand{\ms}{{\mathrm{MS}}}

\newcommand{\kmeans}{{\it K}-means }
\newcommand{\cluster}{\mathcal{C}}
\newcommand{\dmino}{d_{min}^0}

\newcommand{\cM}{\mathcal{M}}
\newcommand{\cF}{\mathcal{F}}

\newcommand{\cCo}{\mathcal{C}}

\newcommand{\vmu}{\mbox{\boldmath$\mu$\unboldmath}}

\newcommand{\mx}{\mathbf{x}}

\newcommand{\beq}{\begin{equation}}
\newcommand{\eeq}{\end{equation}}
\newcommand{\beqa}{\begin{eqnarray}}
\newcommand{\eeqa}{\end{eqnarray}}
\newcommand{\corcol}{\color{black}}

\newcommand{\normalized}{}

\begin{document}

\title{{\it K}-Means Fingerprint Clustering for Low-Complexity Floor Estimation in Indoor Mobile Localization}

\author{\IEEEauthorblockN{Alireza Razavi, Mikko Valkama, and Elena-Simona Lohan}
\IEEEauthorblockA{Department of Electronics and Communications Engineering, Tampere University of Technology, Tampere, Finland\\
email: alireza.razavi@tut.fi, mikko.e.valkama@tut.fi, elena-simona.lohan@tut.fi
}}
\maketitle
%

%

%






\begin{abstract}
Indoor localization in multi-floor buildings is an important research problem. Finding  the correct floor, in a fast and efficient manner, in a shopping mall or an unknown university building  can save the users' search time and can enable a myriad of Location Based Services in the future. One of the most widely spread techniques for floor estimation in multi-floor buildings is the fingerprinting-based localization using  Received Signal Strength (RSS) measurements coming from indoor networks, such as WLAN and BLE (Bluetooth Low Energy). The clear advantage of RSS-based floor estimation is its ease of implementation on a multitude of mobile devices at the Application Programming Interface (API) level, because RSS values are directly accessible through API  interface. However, the downside of a fingerprinting approach, especially for large-scale floor estimation and positioning solutions, is their need to store and transmit a huge amount of fingerprinting data.  The problem becomes more severe when the localization is intended to be done on mobile devices (smart phones, tablets, etc.) which have limited memory, power, and computational resources. An alternative floor estimation method, which has lower complexity and is faster than the fingerprinting is the Weighted Centroid Localization (WCL) method. The trade-off is however paid in terms of a lower accuracy than the one obtained with traditional fingerprinting with Nearest Neighbour (NN) estimates. 
In this paper a novel \kmeans-based method for  floor estimation via fingerprint clustering of WiFi and various other positioning sensor outputs is introduced.  Our method achieves a floor estimation accuracy close to the one with NN fingerprinting, while significantly improves the complexity and the speed of the floor detection algorithm.  The decrease in the database size is achieved through storing and transmitting only the cluster heads (CH's) and their corresponding floor labels. The performance of the proposed methods is evaluated using real-life indoor measurements taken from four multi-floor buildings. The numerical results show that the proposed \kmeans-based method offers an excellent trade-off between the complexity and performance.
\end{abstract}
\begin{IEEEkeywords}
floor estimation, indoor localization, received signal strength (RSS), z-coordinate estimation, fingerprinting localization, clustering, weighted centroid localization.
\end{IEEEkeywords}


\section{Introduction}
\label{sec:intro}
Indoor localization and floor estimation in multi-floor buildings are becoming more and more important in today's wireless world. Being able to achieve accurate ubiquitous localization on hand-held battery-operating mobile devices in both indoor and outdoor environments would open the window to many new Location Based Services (LBS) spanning from person and asset tracking and personal navigation to health remote monitoring and LBS-based social networking. However, despite the fact that outdoor global localization solutions exist nowadays with the help of Global Navigation Satellite Systems (GNSS), global solutions for indoor localization and floor estimation are still hard to find.  

One of the crucial aspects in indoor positioning is to accurately identify the floor where the user is located. False floor estimation could lead  not only to waste of time, for example in Location Based Services dedicated to advertising and shopping assistance, but also to serious injury  or loss of lives in emergency Location Based Services.
 The floor detection or estimation can be either performed in an initial step, before accurate x-y localization, or jointly with the x-y localization \cite{Maneerat, Gansemer}. In our paper we focus on the first case.
 
Typically,  the floor estimation is achieved via fingerprinting approaches with Nearest Neighbor (NN) method, and such approaches can solve the indoor localization problem locally, as Skyhook and Polaris have proved \cite{polaris12}. Such solutions are however expensive and computationally rather expensive to be used on a global scale (e.g. worldwide). In the fingerprinting-based methods \cite{bahl00, honkavirta09, roos02, seco09,  shrestha13}, the location service providers construct a fingerprint database, transfer this database to the Mobile Station (MS), and the MS then computes its location and corresponding floor based on similar fingerprints. The fingerprint databases are typically very large since they do contain Received Signal Strengths (RSS's) coming from various Access Points (APs) and in many points or coordinates within a building. Thus, if a global floor estimation solution would use a fingerprinting approach, the fingerprint database transfered from the server to the MS would include the fingerprints from all essential buildings in the town (or the location area) where the mobile is situated. For example, assuming that i) we hear an average of $30$ APs in each location point inside a building (a location point here refers to the ($x,y,z$) coordinates inside a building where a measurement is done), ii) we take measurements from an average of $600$ location points per building and iii) there are $25$ important buildings (malls, shopping centers, hospitals, airports, ...) in the location area where the mobile was identified by the network, then a total of $495000$ parameters would need to be stored in the database pertaining to that town and transfered to the mobile. The parameters are the fingerprints, namely the $(x,y,z)$ coordinates and the measured RSS values per coordinate (one RSS per each AP heard at that coordinate). In addition, if these parameters are saved with a $32$-bit accuracy, the database size of such a server provider for the particular town of our example would be around $15.86$ Mbits. With average typical cell-edge (coverage) rates in the order of 50-100kbps, especially in legacy networks, the transfer to the mobile would take \emph{several minutes}, which is clearly unacceptable. Moreover, the server would have to deal with hundreds or thousands of user requests for localization, and this could thus easily create a bottleneck also on the server side. 

An alternative floor estimation approach is based only on the known or estimated positions  of the transmitters or APs and some form of trilateration \cite{Huan, Lohan15}. The simplest of such approaches is a weighted centroid approach, described  for example in 
\cite{Lohan15, Liu, Sharma}. While the complexity of such an approach is much less than the one of the NN fingerprinting, their performance leaves a lot to be desired.

Thus to solve the problem of huge databases and to increase the estimation speed, while still achieving high floor detection probabilities, we propose in this paper a novel clustering method, which achieves and outperforms the floor detection probability of NN fingerprinting, but with much lower complexity. The proposed approach is not limited to the WiFi fingerprints, but rather can be applied to the data collected from various other wireless technologies that support RSS measurements, such as cellular data, BLE data, RFID data, etc.

{\bf Related works and the novelty of our work: } Clustering of the fingerprints has been already studied in several papers \cite{kuo07,mo12,swangmuang08,wigren07,youssef03}. In all of these papers, the idea is to divide the fingerprints into several clusters where the size of each cluster is much smaller than the whole set of fingerprints. Then the positioning phase includes two stages: first, the MS observation vector is compared to all CH's, and after finding the most similar cluster, in the second stage the comparisons are done within that cluster. In other words, the whole fingerprinting data is needed for localization. One cannot solely rely on the CH's to perform the positioning task. 
This might not be a problem when the localization task is carried out in the server side as normally servers have powerful processing capability and sufficient power supply, but if the mobile device itself wants to accomplish the positioning task, then the server has to send its whole data to the mobile device which is impractical because of both the transmission of the huge dataset and limited processing capability and power supply on mobile devices \cite{liu07}.

To the best of our knowledge, the method proposed in this paper is the only attempt in clustering the fingerprinting data which needs only the CH's information for $z$-coordinate positioning task and therefore is implementable on mobile devices. A more detailed comparison between our proposed clustering method and the existing clustering methods for localization will be given in Section \ref{sec:comparison} after introducing our method. A numerical comparison is also carried out against the clustering approach proposed in \cite{kuo07}.

{\bf Paper Organization:} In Section \ref{sec:background} the basics of two conventional methods for indoor localization are described. The proposed clustering approach for floor estimation is then introduced in Section \ref{sec:algorithm}. We will also explain the essence of existing clustering algorithms and provide a brief analytical comparison between the complexity of our proposed clustering method and the existing clustering algorithms. In Section \ref{sec:numerical}, we study the performance of the proposed algorithm in terms of probability of floor detection and computational complexity against the two conventional method described in Section \ref{sec:background} as well as an existing clustering approach based on real-life measurements in four different multi-storey buildings. Finally we conclude the paper in Section \ref{sec:conclusion}.

{\bf Mathematical notations:} Throughout this paper vectors and matrices are shown with the small and capital bold letters, respectively. The second norm of a vector is denoted by $\| \cdot \|_2$. For a set $\cM$, the cardinality of the set is shown as $|\cM|$. Equality is denoted by $=$ and definition is denoted by $\triangleq$. For a real number $a$, the smallest integer number bigger than or equal to $a$ is denoted by  $\lceil a \rceil$.

\section{Conventional Approaches for Indoor Localization}
\label{sec:background}

In this section, we shortly describe two conventional methods for indoor localization: the Nearest Neighbor (NN) fingerprinting localization which is a high-complexity but very promising method, and Weighted Centroid Localization (WCL) which is a low-complexity method but with a performance noticeably inferior to fingerprinting approach.

\subsection{Fingerprinting Localization and Problem Statement}
\label{sec:problemStatement}
{\corcol
Consider a localization system equipped with $\Nap$ positioning signals (e.g., RSS values received from APs). During the offline phase, the positioning signals are collected in $\Nap \times 1$ {\it measurements vectors} $\vx_n \defeq [\sx_{n,1},\sx_{n,2},\ldots,\sx_{n,\Nap}]^T,~$$n=1,\ldots,\Nfp$, where $\Nfp$ is the number of fingerprints collected in the building and $\sx_{n,ap}$ is the RSS received from access point $ap$ at $n$-th collected fingerprint. The corresponding known 3-D location of $\vx_n$ is denoted by $\vc_n \defeq [x_n,y_n,z_n]^T,~n=1,\ldots,\Nfp$. In fingerprinting approach, the fingerprints $\{\vx_n,\vc_n\},~n=1,\ldots,\Nfp$ are stored and used for localization purposes. 


Assume that an MS observes a positioning vector $\vx_{\ms}\triangleq[m_{\ms,1},m_{\ms,2},\ldots,m_{\ms,\Nap}]^T$, where $m_{\ms,ap}$ is the  RSS from $ap$-th AP. The basic 1-NN fingerprinting approach estimates the location of the MS as
\beq
\hat{\vc}_{\ms,\fp} = \vc_j,
\eeq
where
\beq
j = \underset{n \in \{1,\ldots,\Nfp\}}{\arg} \min d(\vx_\ms,\vx_n),
\eeq
where $d(\cdot,\cdot)$ is a dissimilarity measure which is determined based on our assumption for noise. For instance if we assume that the noise which deviates the $\vx_\ms$ from $\vx_n$ is i.i.d white Gaussian, then $d(\vx_\ms,\vx_n)$ is simply the squared Euclidean distance between $\vx_\ms$ and $\vx_n$, i.e., 
\beq
d(\vx_\ms,\vx_n) = \| \vx_\ms - \vx_n \|_2^2.
\eeq
In general, fingerprint-based localization approach is a {\it pattern matching} approach \cite{bahl00, roos02, kuo07}, rooted in pattern recognition \cite{duda_book73}, which tries to match the pattern $\vx_\ms$ observed by MS to the examples $\{\vx_n\}_{n=1}^{\Nfp}$ collected in the training data set and chooses the location of the less dissimilar example (fingerprint) as the location of MS. In this regard, each element of measurements vector $\vx_n$ is a {\it feature} of the location $\vc_n$. On the other hand, any measured signal which depends {\it only on the measurement location} (regardless of noise, shadowing and other uncertainties), can be regarded as a {\it feature} of that location and used for localization using fingerprinting scheme. 

The main problem with the fingerprinting approach is the huge amount of data which must be stored by servers and transmitted to the MS to localize itself when $\Nfp$ is a large number. The situation becomes even more severe when the fingerprints are being collected all over the time. If we want to use fingerprinting methods for localizing the mobile device, it can only be done on the server side. Because of the limited processing capabilitiy and power supply on most of the mobile devices, they are not capable of storing and processing of that huge amount of data \cite{liu07} and furthermore transmitting such amount of fingerprinting data from server to the mobile device takes a lot of time, which makes the localization by mobile devices impractical.

\subsection{Weighted Centroid Localization (WCL)}
Weighted centroid localization approach, first proposed for position estimation in wireless sensor networks \cite{wcl07}, is a simple and low-complexity but promising localization approach. The position of the MS in the WCL approach is computed as the weighted average of the positions of APs heard by the MS.  
Denoting the set of all hearable APs by $\mathcal{H}$ and the (known) coordinates of APs by $\vc_{ap} \triangleq (x_{ap},y_{ap},z_{ap}),~ap=1,\ldots,|\mathcal{H}|$, the WCL-based estimate of mobile station coordinates is computed as
\beq
\hat{\vc}_{\ms,wc} = \frac{\sum_{ap \in \mathcal{H}} w_{ap} \vc_{ap}}{\sum_{ap \in \mathcal{H}} w_{ap} },
\label{eq:centroid1}
\eeq  
where $w_{ap}$ are weight functions. To weight shorter distances (nearer APs) more than higher distances, $w_{ap}$ may be chosen as \cite{wcl07}
\beq
w_{ap} = {1}/{(d_{ap})^g},
\label{eq:wcl_weights}
\eeq 
where $d_{ap}$ is the distance between $ap$-th AP and the MS, and degree $g$ is to ensure that remote APs still impact the position estimation \cite{wcl07}.

Since the distances $d_{ap}$ are not readily available, and also since RSS heard from AP $ap$ is inversely proportional to $d_{ap}$,  the weights $w_{ap}$ in (\ref{eq:centroid1}) can be replaced by RSS  to obtain the following RSS-based formula for WCL \cite{weightedcentroidpatent, wcl11, liu12}
\beq
\hat{\vc}_{\ms,wc} = \frac{\sum_{ap \in \mathcal{H}} m_{\ms,ap} \vc_{ap}}{\sum_{ap \in \mathcal{H}} m_{\ms,ap} },
\label{eq:centroid}
\eeq  
where $m_{\ms,ap}$ is the measured RSS of AP number $ap$ by MS.

Equation (\ref{eq:centroid}) can be written independently for each coordinate. For instance, for the height coordinate (which is the coordinate that matters in floor estimation task) we have
\beq
\hat{z}_{\ms,wc} = \frac{\sum_{ap \in \mathcal{H}} m_{\ms,ap} z_{ap}}{\sum_{ap \in \mathcal{H}} m_{\ms,ap} },
\label{eq:centroidz}
\eeq 
The APs deployed in commercial and privately-owned buildings, such as shopping malls or blocks of flats, are typically owned by various owners and thus their locations are not centralized or even known in totality. In industrial and university buildings, the AP location may be known to some extent, but as seen in our measurement campaigns, such information is typically stored in incomplete or inexact form, because it is not considered important from the communication point of view. For these reasons, we assume in our work that the location of APs is estimated based on the available fingerprint data, and not known in advance. Therefore, to be able to use the WCL approach, we have to first estimate the location of APs in the training phase. To estimate them, we can again employ the WCL approach and apply it on collected fingerprinting data to estimate the AP coordinates. Let us denote the set of fingerprint measurements who hear the AP $ap$ by $\Fap$ and the RSSI heard at $n$-th fingerprint from $ap$-th AP by $m_{n,ap}$. Then the coordinates of the AP $ap$ can be calculated as
\beq
\hat{\vc}_{ap} = \frac{\sum_{n \in \Fap} m_{n,ap} \vc_{n}}{\sum_{n \in \Fap} m_{n,ap} },
\label{eq:centroidAPs}
\eeq  
From (\ref{eq:centroidz}) it is clear that for performing the floor estimation task, the only thing that we need to store at the mobile side is $\Nap$ numbers $z_{ap},~{ap=1,\ldots,\Nap}$ which are the $z$-coordinates of Access Points. This makes the WCL approach a promising one from complexity point of view. However, the performance of the method is relatively poor because the model suggested by (\ref{eq:centroid}) is an inaccurate model.


\section{\kmeans algorithm for Fingerprint Clustering}
\label{sec:algorithm}
In this section, we will introduce a novel algorithm for substantially reducing the size of fingerprinting data using \kmeans clustering algorithm. The proposed algorithm achieves a performance very close to the NN fingerprinting approach while decreasing the computational complexity both in terms of search time and memory size needed for storing the offline data. We first briefly describe the basic form of \kmeans clustering and then introduce our floorwise clustering approach. Finally we compare the complexity of the proposed method with existing fingerprinting clustering approaches.
\subsection{\kmeans clustering}
\label{sec:kmeans}

\kmeans clustering \cite{macqueen67} is a vector quantization method for finding the CH's in a set of unlabeled data \cite{hastie09_book}. The \kmeans algorithm aims to partition $N$ $K$-dimensional observation vectors $\{ \mx_1,\ldots,\mx_N \}$ into $\NC \le N$ clusters by iteratively moving the CH's $\{\vmu_k\}_{k=1}^\NC$, which are the representatives of clusters, to minimize the within cluster sum-of-squares
\beq
\sum_{k=1}^{\NC} \sum_{\mx_i \in \cluster_k} \| \mx_i - \vmu_k \|_2^2,
\label{eq:kmeansObjectiveFunction}
\eeq
with respect to $\{\cluster_1,\cluster_2,\ldots,\cluster_{\NC}\}$, where $\cluster_k$ denotes the $k$-th cluster. Minimizing the objective function in (\ref{eq:kmeansObjectiveFunction}) in general is NP-hard \cite{dasgupta09}. But the \kmeans algorithm provides a suboptimal solution to it by alternating the following two steps in each iteration until convergence \cite{hastie09_book}:
\begin{enumerate}
	\item for each CH, identify the subset of data points which are closer to it than any other CH; and
	\item compute the mean of all data points belonging to each CH and take it as the new CH. 
\end{enumerate}
The above-mentioned \kmeans iterations guarantee the convergence to a stationary point of (\ref{eq:kmeansObjectiveFunction}) \cite{selim84}.

%

\subsection{Fingerprint clustering for floor detection}
\label{sec:algorithmFD}
When the goal of localization is to detect the floor, for example in a multistorey building, the clustering can be applied by clustering the fingerprints collected in each floor separately. The number of clusters can be different from one floor to another.

We want to cluster the fingerprints in a building with $\nfloor$ floors with the ultimate goal to use the CH's for floor detection. Assume that the set of all fingerprints is partitioned as $\cF=\cF_1 \cup \ldots \cup \cF_\nfloor$, where $\cF_f$ denotes the set of fingerprints collected in floor $f$. The floorwise fingerprint clustering then can be accomplished by applying \kmeans clustering algorithm to vectors $\normalized{\vx}_n \in \cF_f$ for each $f$, separately. 
In the detection phase, the CH's in all floors are compared to the MS observation vector $\vx_{\ms}$ and the floor of the most similar cluster head is chosen as our estimate of the floor. Mathematically, denoting the set of the CH's in $f$-th floor by $\{\normalized{\vxch}_{f,1},\ldots,\normalized{\vxch}_{f,\NC_{,f}}\}$ where $\NC_{,f}$ is the number of clusters in the $f$-th floor, we have
\beq
\hat{f} = \underset{f \in \{1,\ldots,F\}}{\arg} \min_{f,k} d(\vx_\ms,\normalized{\vxch}_{f,k}).
\eeq

This method is in general referred to as \kmeans classification \cite[Chapter 13]{hastie09_book} in the literature.

{
\subsection{Comparison to the existing algorithms}
\label{sec:comparison}

To further elaborate the the novelty of the proposed algorithm, here we compare it with the existing fingerprint clustering algorithms based on the formulation given above.

We denote the set of all fingerprint measurements $\cM\triangleq\{\vx_1,\ldots,\vx_{\Nfp}\}$ and the set of the corresponding fingerprint coordinates by $\cCo\triangleq\{\vc_1,\ldots,\vc_{\Nfp}\}$. Furthermore, we denote the set of resulting cluster heads by $\tilde{\cM} \triangleq \{\vxch_1,\ldots,\vxch_{\NC}\}$ and the set of measurement vectors in $c$-th cluster by ${\cM}_c,~c=1,\ldots,\NC$. Clearly we have ${\cM}=\cup_{c=1}^\NC {\cM}_c$. The existing fingerprinting clustering methods all include the following three steps: 
\begin{enumerate}
\item[S1] First we find the most similar element of the set $\tilde{\cM}$ to $\vx_{\ms}$. Denote this element by $\vxch_{\hat{c}}$.
\item[S2] Then we find the most similar element of the set ${\cM}_{\hat{c}}$ to $\vx_{\ms}$. Denote this element by $\vx_{\hat{j}}$. Notice that $\vx_{\hat{j}} \in \cM$.
\item[S3] Finally, our estimate for the coordinate of MS is $\vc_{\hat{j}} \in \cCo$.
\end{enumerate}
The main problem of these methods, which has been solved in our proposed algorithm, is that in order to carry out step {S2}, we have to save all the fingerprints. The only advantage of these clustering algorithms over the ordinary fingerprint matching methods is that after finding the most matched cluster head, the search is limited within only that cluster instead of the whole data base. But since it is not known in advance what cluster will be chosen later in step {S1}, these methods still need to save all of the cluster sets $\cM_c,~c=1,\ldots,\NC$ (whose union is equal to the whole fingerprinting database), the $z$-coordinates of all fingerprints, and the set of the cluster heads $\tilde{\cM}$, which altogether are of size $(\Nap+1) \Nfp + \Nap \NC$.

In our method, on the other hand, we do not need to save $\cM$ and $\cCo$. We are able to estimate the floor only from the cluster heads and their corresponding floor labels which is only of size $(\Nap+1)\NC$. 
In Section \ref{sec:numerical} we will provide a numerical comparison of our method against one of the existing clustering approaches as well as the two conventional floor estimation methods explained in Section \ref{sec:background}.
}

\section{Measurement Results and Analysis}
\label{sec:numerical}

In this section, we study the performance of the proposed clustering algorithm by selected real-life measurement examples.

\subsection{Measurement set-up}


The numerical examples here are based on real-life WLAN data collected in four multi-storey buildings. The first building is a four-floor university building (Univ-1), the second building is a three-floor university building (Univ-2), the third building is a six-floor shopping mall (Mall), and finally the fourth building is a four-floor office building (Office).


We have two sets of data for each building. The first data set includes the fingerprints collected in the building which is used for training. This is the data set to which the fingerprint clustering is applied and afterwards we only use the CH's for floor detection purposes. We refer to this data set as {\it fingerprinting data}. The second set has been collected along several different tracks in each building, where each track includes tens to hundreds of data points. This data set will be used for examining the performance of the proposed algorithm and is referred to as {\it test data}. 

{\corcol The measurement points for collecting fingerprinting data and test data in Univ-1 and Univ-2, are illustrated in Figures \ref{tieto_map} and \ref{sahko_map}, respectively. Figure \ref{powermap} shows the power map of a selected AP, namely AP number 4, in the first floor of Univ-1. This is also the floor where AP 4 is situated. The red points have the highest RSS values which are in fact points closest to the physical location of the AP and the blue points have the lowest RSS values and are the points farthest from the AP.}

A summary of relevant technical details in each building including the number of floors $N_{fl}$, the size of fingerprinting data determined by $\Nfp$, the size of test data determined by the number of test points $N_t$, and the number of Access Points $\Nap$ heard are shown in Table \ref{table1}.

\begin{figure}[t!]
\begin{center}
\includegraphics[width=0.5\textwidth]{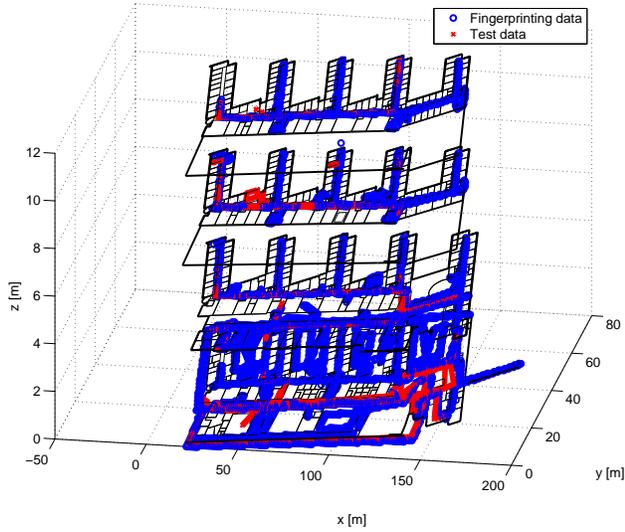}
\caption{Measurement points for collecting fingerprinting data and test data in Univ-1.}
\label{tieto_map}
\end{center}
\end{figure}

\begin{figure}[h]
\begin{center}
\includegraphics[width=0.5\textwidth,height=0.3\textwidth]{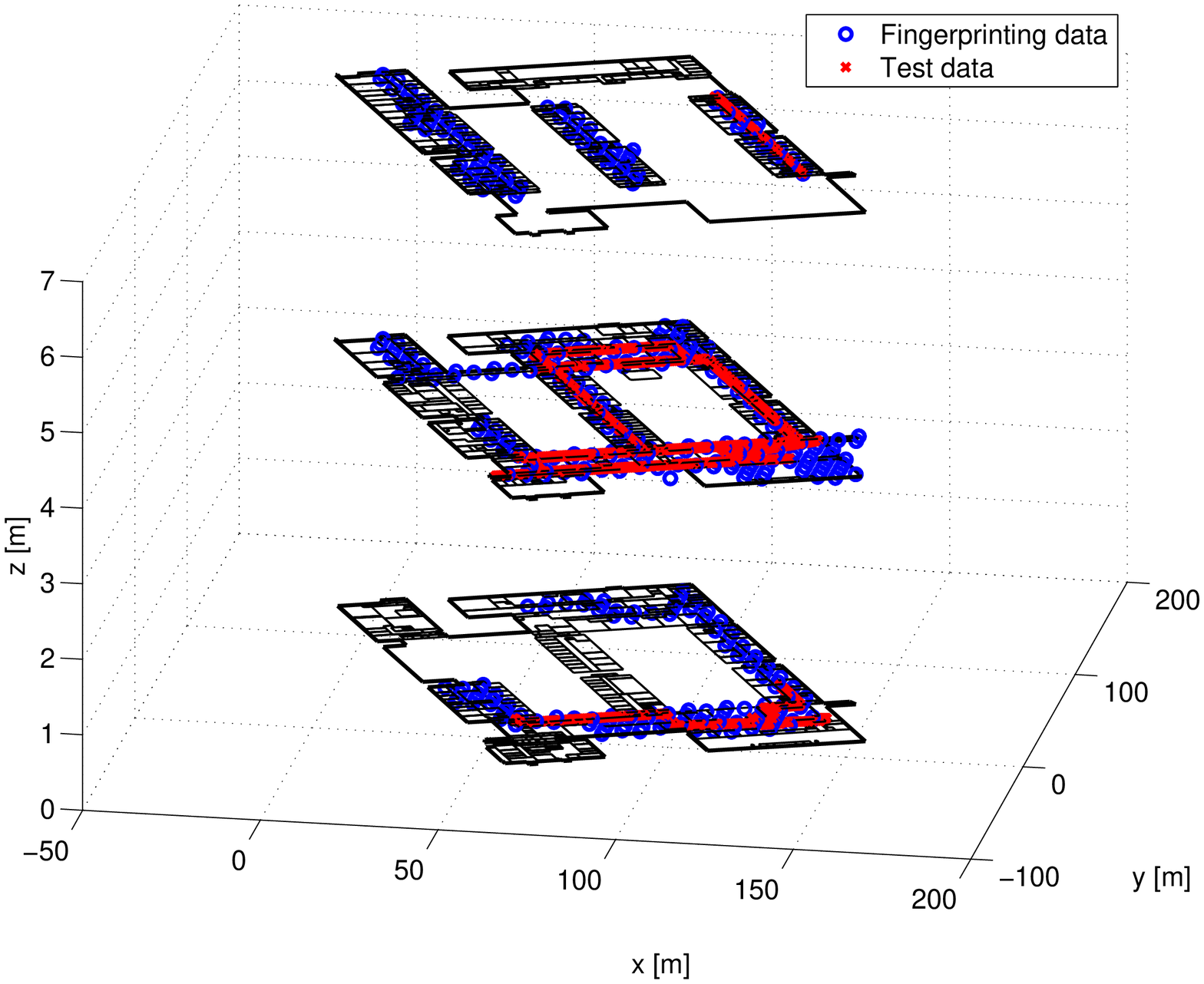}
\caption{Measurement points for collecting fingerprinting data and test data in Univ-2.}
\label{sahko_map}
\end{center}
\end{figure}

\begin{figure}[h]
\begin{center}
\includegraphics[width=0.5\textwidth,height=0.3\textwidth]{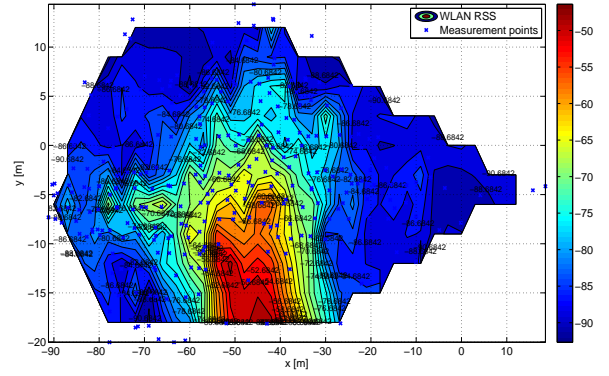}
\caption{Power map (the map of RSS values) of the first floor of Univ-1 for AP number 4. The figure shows how the received signal strength (RSS) from AP number 4 is distributed over the floor.}
\label{powermap}
\end{center}
\end{figure}

\subsection{Numerical study of four methods for floor estimation}

As mentioned in Section \ref{sec:algorithmFD}, we apply the K-means clustering algorithms to the fingerprints in each floor separately. This task can be done on the server side when the computational resources are powerful enough to apply the clustering algorithm to a possibly huge amount of fingerprinting data. The server then sends the computed cluster heads together with the floor label of each cluster head to the mobile device to be used for positioning in the online phase. In other words, the data needed in the mobile side for performing the floor detection task is only limited to the cluster heads and their corresponding floor labels. We remark that if we want to use the ordinary fingerprinting method, the server would need to send the entire fingerprinting data to the mobile device. Therefore by using the clustering algorithm, we will achieve a considerable reduction in the size of data needed to be transmitted to the mobile device for floor detection. The size reduction provides two benefits in the mobile side: first, the memory size required for storing the data (cluster heads) is less than the ordinary fingerprinting, and second, the complexity of the search for finding the most similar cluster head decreases significantly.

\begin{table}[t!]
\centering
\caption{Number of floors $N_{fl}$, number of training samples (fingerprinting data) $\Nfp$, number of test samples $N_t$, and number of Access Points $\Nap$ in each of the four buildings. }
\begin{tabular}{|l|c|c|c|c|}
\hline
Building      & $N_{fl}$& $\Nfp$ & $N_{t}$ & $\Nap$  \\ \hline \hline
Univ-1 & 4 &     16080     &     6796     &    509     \\ \hline
Univ-2 & 3 &     9923     &     2301     &    489      \\ \hline
Mall        &6 &     1633     &      3503    &     468     \\ \hline
Office      & 4 &     354     &    3873      &    1103      \\ \hline
\end{tabular}
\label{table1}
\end{table}


\begin{table*}[t!]
\centering
\caption{Probability of Floor Detection for each method.}
\begin{tabular}{|c||c||c||c|c|c||c|c|c|}
\hline
\multirow{2}{*}{Building} & \multirow{2}{*}{1-NN} & \multirow{2}{*}{WCL} & \multicolumn{3}{c||}{\cite{kuo07}} & \multicolumn{3}{c|}{Proposed Floorwise  Clustering} \\ \cline{4-9} 
                          &                       &                      & $\rho=0.01$            & $\rho=0.05$           & $\rho= 0.1$         & $\rho=0.01$            & $\rho=0.05$           & $\rho= 0.1$                \\ \hline \hline
Univ-1            & 0.8868                 & 0.7016          & 0.7076 &0.7088  &      0.7056      & 0.8973    & 0.8996    & 0.9036             \\ \hline
Univ-2           & 0.9944                 & 0.6784      & 0.9831 & 0.9904 &        0.9904        &  0.9739   & 0.9930 &   0.9870             \\ \hline
Mall                         & 0.9255                  & 0.6041      & 0.8927 & 0.9146 &    0.9212            & 0.8401   & 0.9055  &  0.9366               \\ \hline
Office                        & 0.8084                     & 0.7033         & 0.8012 &  0.7896 & 0.8066            & 0.7088  &  0.7981  &  0.8043             \\ \hline
\end{tabular}
\vskip0.3cm
\label{table:pd}
\end{table*}

\begin{table*}[t]
\centering
\caption{Elapsed time (in seconds) for floor estimation using each method.}
\begin{tabular}{|c||c||c||c|c|c||c|c|c|}
\hline
\multirow{2}{*}{Building} & \multirow{2}{*}{1-NN} & \multirow{2}{*}{WCL} & \multicolumn{3}{c||}{\cite{kuo07}} & \multicolumn{3}{c|}{Proposed Floorwise  Clustering} \\ \cline{4-9} 
                          &                       &                      & $\rho=0.01$            & $\rho=0.05$           & $\rho= 0.1$         & $\rho=0.01$            & $\rho=0.05$           & $\rho= 0.1$                \\ \hline \hline           
Univ-1             & 32.62                 & 0.066               & 0.67& 1.229&     2.769  & 0.35            & 1.25            & 3.92                    \\ \hline
Univ-2           & 5.935                 & 0.026                 & 0.228 & 0.284 &  0.511   & 0.095          & 0.258          & 0.772                     \\ \hline
Mall                        & 0.832                   & 0.031              &0.171 & 0.155 &     0.195   & 0.069               & 0.100               & 0.180                        \\ \hline
Office                         & 0.366                     & 0.035             & 0.218 & 0.160 &    0.162     & 0.084              & 0..095               & 0.137                           \\ \hline
\end{tabular}
\vskip0.3cm
\label{table:time}
\end{table*}

\begin{table*}[t]
\centering
\caption{The size of data (in Kilo Bytes) needed at the mobile end for performing the floor detection task.}
\begin{tabular}{|c||c||c||c|c|c||c|c|c|}
\hline
\multirow{2}{*}{Building} & \multirow{2}{*}{1-NN} & \multirow{2}{*}{WCL} & \multicolumn{3}{c||}{\cite{kuo07}} & \multicolumn{3}{c|}{Proposed Floorwise  Clustering} \\ \cline{4-9} 
                          &                       &                      & $\rho=0.01$            & $\rho=0.05$           & $\rho= 0.1$         & $\rho=0.01$            & $\rho=0.05$           & $\rho= 0.1$                \\ \hline \hline
Univ-1            	& 864                 & 0.412            & 942 & 1100 &  1300        & 78            & 319          & 692     \\ \hline
Univ-2          	& 553                 & 0.389               &602&709&811       & 53            & 203           & 451         \\ \hline
Mall                         	& 315                 & 0.534            & 332 & 356 &   381       & 20    	& 61    	&123         \\ \hline
Office                         	& 262                 & 0.663           & 274 & 295 &     315      & 20 		&  53 		& 106 	        \\ \hline
\end{tabular}
\label{table:size}
\end{table*}

Tables \ref{table:pd}, \ref{table:time}, and \ref{table:size} show, respectively, the probability of floor detection, the time for floor estimation in online phase\footnote{The time is the running time of floor detection algorithm on a computer with 2.2 GHz Intel Core i7 CPU and 16 GB of random access memory (RAM).}, and the size of data needed on the mobile device for four different methods: ordinary 1-Nearest Neighbor (1-NN) fingerprint positioning, the Weighted centroid localization (WCL) method, the clustering approach proposed in \cite{kuo07}, and our proposed floorwise clustering algorithm. For the method proposed in \cite{kuo07} and also for our approach, the clustering is implemented for three different {\it clustering ratios} $\rho \in \{0.01,0.05,0.1\}$, where the clustering ratio here has the following relationship with the number of cluster heads in $f$-th floor, $N_{\mathrm{ch},f}$, and the total number of fingerprints in $f$-th floor, $N_{\mathrm{fp},f}$:
\beq
N_{\mathrm{ch},f} = \lceil \rho \times  N_{\mathrm{fp},f} \rceil.
\eeq 

As it can be seen from Table \ref{table:pd}, even with the very small clustering ratio of $\rho=0.01$, the performance of the proposed floorwise clustering approach is very close to the 1-NN fingerprinting approach for the first two buildings. The performance for the next two buildings, namely Mall and Office, however degrades which is mainly because of relatively small number of fingerprints collected in these two buildings (see table \ref{table1}). The performance is clearly superior to the WCL approach. The performance of \cite{kuo07} is similar to the 1-NN fingerprinting approach except for the first building (which has the highest number of fingerprints) where its performance degrades significantly. This is because when the number of fingerprints is very large, it is more likely that the method in \cite{kuo07} puts the fingerprints from different floors to the same cluster which may eventually results in an erroneous estimate for the floor.

For higher values of $\rho$ on the other hand, the performance of our proposed method becomes very similar to 1-NN fingerprinting. For some points it even slightly outperforms the 1-NN fingerprinting which is because like any other compression algorithm the clustering approach provides some denoising too.

From Table \ref{table:time} we can see that the time needed for computing the estimated floor using our proposed clustering approaches is much faster than that of ordinary fingerprinting approach. It can be seen that the proposed method is also faster than the clustering approach of \cite{kuo07} and is only inferior to WCL which has a rather poor floor estimation performance. 

From Table \ref{table:size}, it can be seen that the size of data needed at the mobile side for our method is noticeably smaller than those of 1-NN fingerprinting approach and the clustering approach of \cite{kuo07}, especially for small clustering ratios. The method of \cite{kuo07} needs the highest memory size that as explained in Section \ref{sec:comparison}, is because it has to store the cluster heads in addition to all fingerprints. 

{\corcol Finally, Figure \ref{fig3d} illustrates an overall comparison between the four mentioned methods based on the three performance metrics of i) the elapsed time for computing the floor estimation, ii) the data size needed in the mobile side, and iii) the probability of floor detection. The clustering ratio for the clustering method of \cite{kuo07} and our proposed clustering method is $\rho=0.01$. The colors are to distinguish between the methods and the four vertices of each pyramid are corresponding to the four buildings under study in the experiment. As it can be seen the proposed method (red pyramid) is much closer to the origin of the horizontal plane than \cite{kuo07} (blue pyramid) and 1-NN (cyan pyramid) which means it has a much lower complexity than them. It is not as close as the WCL (magenta pyramid), but instead it delivers a much better floor detection probability than WCL.
Thus, overall, the proposed method provides excellent floor detection performance while being able to reduce the computing and data transfer complexities in a substantial manner.}
\begin{figure}
\begin{center}
\includegraphics[width=0.5\textwidth,height=0.33\textwidth]{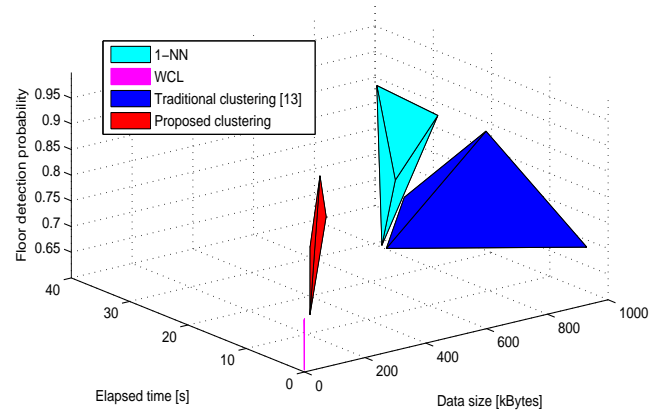}
\caption{Comparison between the four methods based on the 3 performance metrics, namely elapsed time, data size, and floor detection probability. The clustering ratio for \cite{kuo07} and our proposed method is $\rho=0.01$.}
\label{fig3d}
\end{center}
\end{figure}

\section{Conclusion}
\label{sec:conclusion}

With the goal of substantially reducing the size of fingerprinting data needed for storage and transmission in floor estimation, a method for clustering fingerprints using \kmeans algorithm was proposed. The proposed technique applies the clustering algorithm floorwise and keeps and transmits only the cluster heads together with their corresponding floor labels for floor estimation.
The performance of the proposed methods was evaluated with comprehensive real-life indoor measurements. The obtained results show that while the proposed method delivers a significant enhancement in the speed of floor estimation algorithm and a substantial reduction in the size of fingerprint database needed at the mobile device, its localization performance is in par with the conventional fingerprinting approach which uses all the data for accomplishing localization task.    



\begin{thebibliography}{10}

\bibitem{Maneerat} 
K. Maneerat and C. Prommak, \newblock ``An Enhanced Floor
 Estimation Algorithm for Indoor
Wireless localization systems using confidence
interval approach'', \newblock In {\em International Journal of Computer, 
Control, Quantum and Information Engineering} Vol:8, No:7, 2014



\bibitem{Gansemer} S. Gansemer, U. Grossmann, and S. Hakobyan, 
"\newblock ``RSSI-based euclidean
distance algorithm for indoor positioning adapted for the use in
dynamically changing WLAN environments and multi-level buildings'',"
\newblock In {\em 2010 Int. Conf. Indoor Positioning and Indoor Navigation}, pp. 1–6.


\bibitem{Huan}
H. H. Liu; Y. N. Yang, 
\newblock ``WiFi-based indoor positioning for multi-floor Environment," 
\newblock {\em TENCON 2011 - 2011 IEEE Region 10 Conference}, pp.597,601, 21-24 Nov. 2011
doi: 10.1109/TENCON.2011.6129175

\bibitem{Lohan15}
E.S. Lohan, J. Talvitie, P. Figueiredo e Silva, H. Nurminen, S. Ali-L\"{o}ytty, R. Pich{\'e}, "
\newblock ``Received signal strength models for WLAN and BLE-based 
indoor positioning in multi-floor buildings''", 
\newblock In {\em Proc. of IEEE ICL GNSS}, Sweden, Jun 2015 


\bibitem{Liu}
A. Liu, T. Robison, 
\newblock ``Cost and accuracy: factors concerning
various indoor location estimation methods'', 
\newblock 2004, https://courses.cs.washington.edu/courses/
cse561/04au/projects/papers/A.Liu-Robison.pdf, (accessed Jun 2015)


\bibitem{Sharma}
N. K. Sharma, 
\newblock ``A weighted center of mass based trilateration 
approach for locating wireless devices in indoor environment'', 
\newblock  In {\em Proceedings of the 4th ACM international workshop on Mobility 
management and wireless access (MobiWac '06)}, New York, NY, USA, pp. 112-115, 2006

\bibitem{bahl00}
P. Bahl and V.~N. Padmanabhan.
\newblock ``Radar: An in-building RF-based user location and tracking system'',
\newblock In {\em INFOCOM 2000. Nineteenth Annual Joint Conference of the IEEE
  Computer and Communications Societies. Proceedings. IEEE}, volume~2, pages
  775--784,  2000.

\bibitem{wcl07}
J.~Blumenthal, R.~Grossmann, F.~Golatowski, and D.~Timmermann.
\newblock ``Weighted centroid localization in zigbee-based sensor networks''.
\newblock In {\em Intelligent Signal Processing, 2007. WISP 2007. IEEE
  International Symposium on}, pages 1--6. IEEE, 2007.

\bibitem{dasgupta09}
S.~Dasgupta and Y.~Freund.
\newblock ``Random projection trees for vector quantization''.
\newblock {\em Information Theory, IEEE Transactions on}, 55(7):3229--3242,
  July 2009.

\bibitem{duda_book73}
R.~O. Duda, P.~E. Hart, et~al.
\newblock {\em Pattern classification and scene analysis}, volume~3.
\newblock Wiley New York, 1973.

\bibitem{hastie09_book}
T. Hastie, R. Tibshirani, and J. Friedman.
\newblock {\em The elements of statistical learning}, volume~2.
\newblock Springer.

\bibitem{honkavirta09}
V. Honkavirta, T. Perala, S. Ali-L\"{o}ytty, and R. Pich{\'e}.
\newblock ``A comparative survey of WLAN location fingerprinting methods.''
\newblock In {\em Positioning, Navigation and Communication, 2009. WPNC 2009.
  6th Workshop on}, pages 243--251. IEEE, 2009.

\bibitem{kuo07}
S. P. Kuo, B. J. Wu, W. C. Peng, and Y. C. Tseng.
\newblock ``Cluster-enhanced techniques for pattern-matching localization
  systems''.
\newblock In {\em Mobile Adhoc and Sensor Systems, 2007. MASS 2007. IEEE
  Internatonal Conference on}, pages 1--9. IEEE, 2007.

\bibitem{liu12}
Y.~Liu, X.~Yi, and Y.~He.
\newblock ``A novel centroid localization for wireless sensor networks,''
\newblock {\em International Journal of Distributed Sensor Networks}, 2012.

\bibitem{macqueen67}
J. MacQueen et~al.
\newblock ``Some methods for classification and analysis of multivariate
  observations.''
\newblock In {\em Proceedings of the fifth Berkeley symposium on mathematical
  statistics and probability}, volume~1, page~14. California, USA, 1967.

\bibitem{mo12}
Y. Mo, Z. Cao, and B. Wang.
\newblock ``Occurrence-based fingerprint clustering for fast pattern-matching
  location determination''.
\newblock {\em Communications Letters, IEEE}, 16(12), 2012.

\bibitem{weightedcentroidpatent}
J.~Murray and B.~Tarlow.
\newblock Wi-Fi position fix.
\newblock Europan patent application EP 2574954A1, 2013.

\bibitem{roos02}
T. Roos, P. Myllym{\"a}ki, H. Tirri, P. Misikangas, and J.
  Siev{\"a}nen.
\newblock ``A probabilistic approach to WLAN user location estimation.''
\newblock {\em International Journal of Wireless Information Networks},
  9(3):155--164, 2002.

\bibitem{seco09}
F. Seco, A.~R. Jim{\'e}nez, C. Prieto, J. Roa, and K.
  Koutsou.
\newblock ``A survey of mathematical methods for indoor localization.''

\bibitem{selim84}
S.~Z. Selim and M.~A. Ismail.
\newblock ``K-means-type algorithms: a generalized convergence theorem and
  characterization of local optimality.''
\newblock {\em Pattern Analysis and Machine Intelligence, IEEE Transactions
  on}, (1):81--87, 1984.

\bibitem{shrestha13}
S. Shrestha, J. Talvitie, and E. S. Lohan.
\newblock ``On the fingerprints dynamics in WLAN indoor localization.''
\newblock In {\em ITS Telecommunications (ITST), 2013 13th International
  Conference on}, pages 122--126. IEEE, 2013.

\bibitem{swangmuang08}
N.~Swangmuang and P.~Krishnamurthy.
\newblock ``On clustering rss fingerprints for improving scalability of
  performance prediction of indoor positioning systems.''
\newblock In {\em Proceedings of the first ACM international workshop on Mobile
  entity localization and tracking in GPS-less environments}, pages 61--66.
  ACM, 2008.

\bibitem{wcl11}
J.~Wang, P.~Urriza, Y.~Han, and D.~Cabric.
\newblock ``Weighted centroid localization algorithm: theoretical analysis and
  distributed implementation.''
\newblock {\em Wireless Communications, IEEE Transactions on},
  10(10):3403--3413, 2011.

\bibitem{wigren07}
T. Wigren.
\newblock ``Adaptive enhanced cell-ID fingerprinting localization by clustering
  of precise position measurements.''
\newblock {\em Vehicular Technology, IEEE Transactions on}, 56(5):3199--3209,
  2007.


\bibitem{youssef03}
M.~A. Youssef, A. Agrawala, and A.~U.~Shankar.
\newblock ``WLAN location determination via clustering and probability
  distributions.''
\newblock In {\em Pervasive Computing and Communications, 2003.(PerCom 2003).
  Proceedings of the First IEEE International Conference on}, pages 143--150.
  IEEE, 2003.

\bibitem{polaris12}
N.~Zhang and J.~Feng.
\newblock ``Polaris : A fingerprint-based localization system over wireless
  networks.''
\newblock Book Chapter, Lecture Notes in Computer Science, Springer, 2012.

\bibitem{liu07}
H. Liu, H. Darabi, and P. Banerjee, \newblock ``Survey of wireless indoor positioning techniques and systems,'' \newblock {\em Systems, Man, and Cybernetics, Part C: Applications and Reviews, IEEE Transactions on}, 37(6): 1067-1080, Nov. 2007.

\end{thebibliography}

\end{document}